\documentclass[a4paper,12pt]{spieman}  
\usepackage{amsmath,amsfonts,amssymb}
\usepackage{graphicx}
\usepackage{setspace}
\usepackage{tocloft}
\usepackage{sidecap,booktabs}
\usepackage{enumitem}
\usepackage{mathtools}

\title{Quantitative photoacoustic oximetry imaging by 
multiple illumination learned spectral decoloring}

\author[ a]{Thomas Kirchner}
\author[ a,*]{Martin Frenz}
\affil[a ]{Biomedical Photonics, Institute of Applied Physics, University of Bern, Bern, Switzerland}

\cftpagenumbersoff{figure}
\cftpagenumbersoff{table} 
\begin{document} 
\maketitle

\begin{abstract}

\noindent \textbf{Significance:} Quantitative measurement of blood oxygen saturation (sO$_2$) with photoacoustic (PA) imaging is one of the most sought after goals of quantitative PA imaging research due to its wide range of biomedical applications.

\noindent \textbf{Aim:} A method for accurate and applicable real-time quantification of local sO$_2$ with PA imaging.

\noindent \textbf{Approach:} We combine multiple illumination (MI) sensing with learned spectral decoloring (LSD); training on Monte Carlo simulations of spectrally colored absorbed energy spectra, in order to apply the trained models to real PA measurements. We validate our combined MI-LSD method on a highly reliable, reproducible and easily scalable phantom model, based on copper and nickel sulfate solutions.

\noindent \textbf{Results:} With this sulfate model we see a consistently high estimation accuracy using MI-LSD, with median absolute estimation errors of $2.5$ to $4.5$ percentage points. We further find fewer outliers in MI-LSD estimates compared to LSD. Random forest regressors outperform previously reported neural network approaches.

\noindent \textbf{Conclusions:} Random forest based MI-LSD is a promising method for accurate quantitative PA oximetry imaging.
\end{abstract}


{\noindent \footnotesize\textbf{*} \linkable{frenz@iap.unibe.ch} }

\begin{spacing}{1}

\section{Introduction}
\label{sec:intro}
A robust and accurate quantitative measurement of blood oxygen saturation (sO$_2$) with photoacoustic (PA) imaging, also called optoacoustic imaging, is one of the most sought after goals of quantitative PA imaging (qPAI) research due to its wide range of immediate applications. Usually qPAI research aims to achieve an absolute quantification of optical properties, like the absorption coefficient $\mu_\textrm{a}$, from measured PA signals $S(\boldsymbol{d},t)$ recorded at times $t$ at detector position $\boldsymbol{d}$ \cite{cox2012quantitative, cox2006two}. In brief, such a quantification of $\mu_\textrm{a}$ encompasses a solution of two ill posed inverse problems. (1) The acoustic inverse problem from $S(\boldsymbol{d},t)$ to an initial pressure spatial distribution $p_0(\boldsymbol{x})$. And (2) the optical inverse problem from $H(\boldsymbol{x}_0) =  p_0(\boldsymbol{x}_0) / \Gamma(\boldsymbol{x}_0) =  \phi(\boldsymbol{x}_0, \mu_\textrm{a}(\boldsymbol{x}), \mu_\textrm{s}'(\boldsymbol{x})) \cdot \mu_\textrm{a}(\boldsymbol{x}_0)$ to $\mu_\textrm{a}(\boldsymbol{x}_0)$, at a location $\boldsymbol{x}_0$, with the Grüneisen parameter $\Gamma$ and the reduced scattering coefficient $\mu_\textrm{s}'$. There the fluence $\phi$ is dependent on unknowns like the absorption and scattering in the tissue surrounding $\boldsymbol{x}_0$. qPAI methods either depend on model-based inversion  \cite{cox2006two,tzoumas2016eigenspectra,perekatova2017fluence,glatz2011blind,ulrich2020reliability,ulrich2019spectral} or data-driven approaches \cite{kirchner2018context,luke2019net,cai2018end,yang2019quantitative,durairaj2020unsupervised,bench2020toward}. These approaches often perform well \emph{in silico} but struggle with the translation to real measurements in either phantoms or \emph{in vivo}.

In PA imaging, sO$_2$ estimations are derived from multispectral PA measurements by first performing an acoustic reconstruction yielding images of the PA signal
\begin{equation}
    S(\boldsymbol{x}_0, \lambda) = \Gamma(\boldsymbol{x_0}) \cdot A(\boldsymbol{x}_0) \cdot \phi(\boldsymbol{x}_0, \mu_\textrm{a}(\boldsymbol{x}, \lambda), \mu_\textrm{s}'(\boldsymbol{x}, \lambda)) \cdot \mu_\textrm{a}(\boldsymbol{x}_0)
\end{equation} 
for each measured wavelength $\lambda$, with $A(\boldsymbol{x}_0)$ being an unknown spatially varying factor introduced by the imperfectly solved acoustic ill-posed inverse problem (i.e.\ image reconstruction from data with limited frequency bandwidth and a limited probe aperture). Using a linear image reconstruction, the acoustic inverse problem can be assumed as wavelength independent. The spectral coloring \cite{cox2012quantitative} due to the wavelength dependent fluence variation causes the dominant distortion in any sO$_2$ estimation made from multispectral signal stacks $S(\boldsymbol{x}, \boldsymbol{\lambda})$.
This spectral coloring of PA signals needs to be corrected for in order to yield accurate quantitative estimates of sO$_2$.
To address this need, we combine two approaches to quantitative PA imaging of sO$_2$. (1) Multiple illumination (MI) sensing \cite{held2016multiple} -- a method in which a sequence of PA measurements is acquired with a sequence of illuminations at different positions. Usually, effective attenuation of the illumination is then estimated with diffusion theory and then used for correcting spectral coloring. (2) Learned spectral decoloring (LSD)\cite{grohl2021learned} -- a data science method in which a machine learning algorithm is trained on Monte Carlo simulations of spectrally colored multispectral PA measurements in order to decolor real measurements. 

Both these methods can yield promising results on their own but still suffer from a range of constraints. I.e.\ MI sensing implementations \cite{shao2011estimating} typically assume and use point illuminations which enables the use of closed-form solutions of the diffusion approximation of light propagation\cite{held2016multiple}, but limit SNR due to the laser safety limit for skin \cite{LaserSafetyANSI2005}. The resulting long acquisition times make this method difficult to translate to realistic macroscopic applications \cite{kim2020correction}. Furthermore, MI sensing so far has theoretical limits in highly inhomogeneous scenes due to its reliance on the diffusion approximation. MI sensing implementations usually aim to estimate absolute values of $\mu_\textrm{a}$, which goes beyond what is needed for an estimation of sO$_2$.
LSD \cite{grohl2021learned, grohl2019estimation} and similar spectral approaches \cite{tzoumas_eigenspectra_2016} currently yield accurate \emph{in silico} estimations and plausible initial results in highly constrained settings but they have insufficient input to robustly generalize these results over diverse geometries and applications.
Both MI sensing and LSD are not yet thoroughly validated partially due to a lack of stable and reliable sO$_2$ phantoms.

Even though substantial progress has been made in dynamic blood flow phantoms for photoacoustic imaging validation, these blood or red blood cell suspension phantoms require extensive fine tuning and even then yield reference values with limited accuracy \cite{vogt2019photoacoustic}. At best a reference measurement of $2 - 4$\,\% is achievable with state of the art blood flow phantoms. \cite{ laufer2005vitro, mitcham2017}

Rather than implement such a sO$_2$ flow phantom we used copper and nickel sulfate solutions in a relative copper sulfate model similar to work by Buchmann et al.\cite{buchmann2020quantitative} to mimic absorption spectra of differently oxygenated blood. This allowed a reliable sub $1$\,\% error in our ground truth and allowed us to rapidly manufacture stable and highly reproducible phantoms with wide variations in optical properties in order to generate high quality test sets for spectral decoloring methods.

\section{Materials and methods}
We investigated a method combining learned spectral decoloring (LSD) and multiple illumination (MI) measurements. To that effect we

\begin{enumerate}
    \item Developed a system to perform real-time multiple illumination multispectral PA imaging.
    \item Implemented modified LSD machine learning algorithms using MI.
    \item Used these algorithms to train on \emph{in silico} data from Monte Carlo optical forward simulations with a relative copper sulfate model.
    \item Validated and tested machine learning models trained only on \emph{in silico} data on comprehensive phantom measurements using that copper and nickel sulfate based sO$_2$ model.
\end{enumerate} 

\subsection{Multiple illumination photoacoustic imaging}
\begin{SCfigure}[][hbt]
\centering
\begin{tabular}{c}
\includegraphics{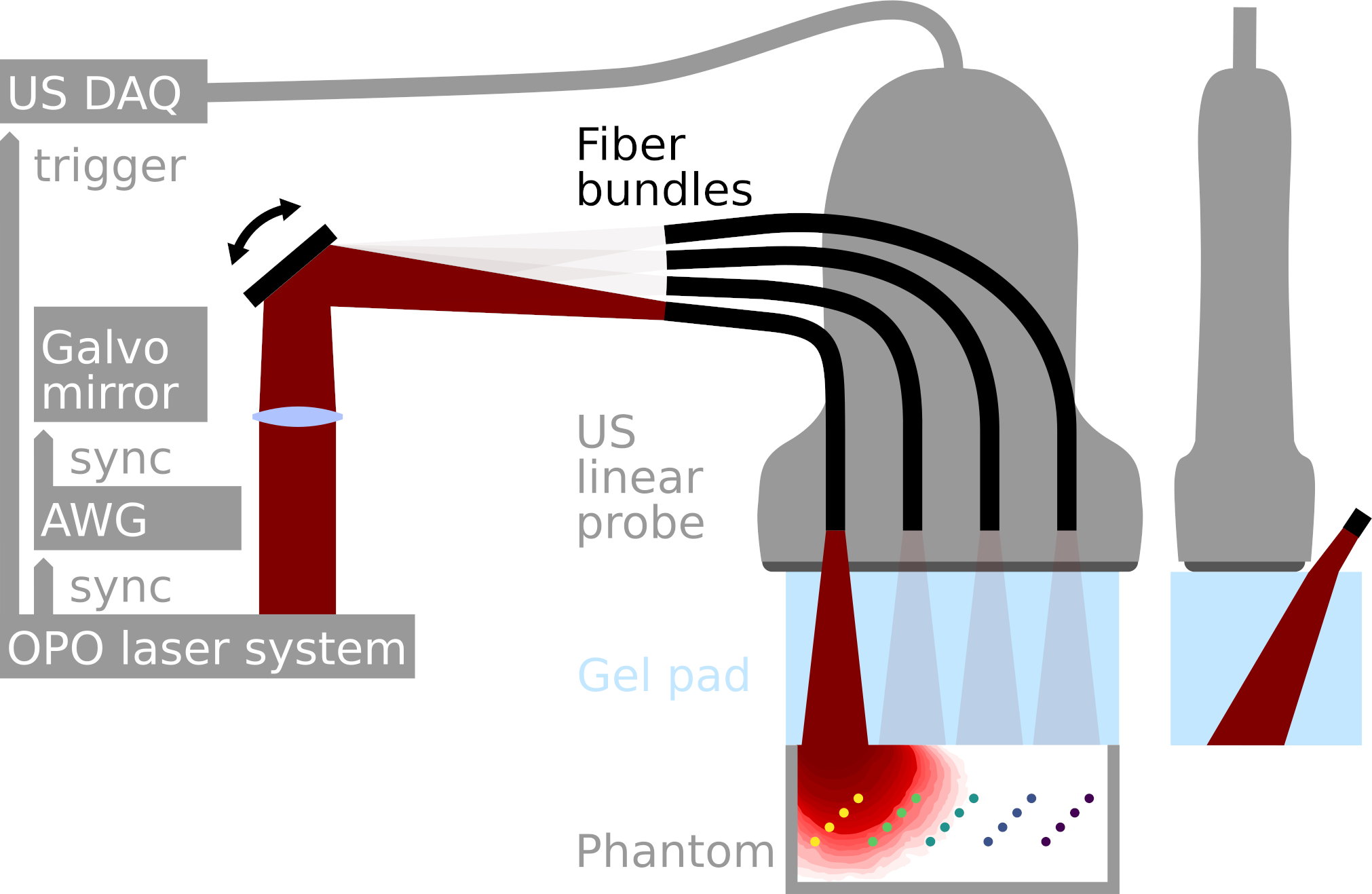}
\end{tabular}
\caption[Multiple illumination (MI) photoacoustic imaging setup.]{\label{fig:setup} Multiple illumination (MI) photoacoustic (PA) imaging setup. Illumination via fast tunable optical parametric oscillator (OPO) laser sequentially illuminating fiber bundles using a galvo mirror system driven by an arbitrary waveform generator (AWG). PA signals were measured with a linear array ultrasound (US) probe and recorded by a 64 channel US data acquisition (DAQ) system. An US gel pad is used to allow for in plane illumination.}
\end{SCfigure}

Our MI PA imaging setup is illustrated in figure\,\ref{fig:setup}. It uses a fast wavelength-tunable optical parametric oscillator (OPO) laser system (prototype SpitLight, InnoLas Laser GmbH, Krailling, Germany) with five nanosecond pulse duration and 100\,Hz pulse repetition frequency. The laser pulses were sequentially coupled into four high power fiber bundles (FiberOptic P.+P.\ AG, Spreitenbach, Switzerland) with NA\ 0.22 fibers, each bundle with a 2\,mm diameter. This was achieved using a galvo mirror system (GVS011/M, Thorlabs Inc., Newton, USA) driven by an arbitrary waveform generator (TG5011, Aim-TTi, Cambridgeshire, United Kingdom), which was synchronized with the laser system. The fiber bundle output sides were arranged in a line array with 8\,mm spacing. The illumination pulses were attenuated to have a maximum energy of 10\,mJ per pulse at the fiber output. To comply with ANSI Safety limits \cite{LaserSafetyANSI2005,Wang2006} the beams are widened to 7\,mm full width at half maximum (FWHM) at the tissue or phantom surface. Illumination and acoustic detection ensues through 18\,mm thick Ultrasound gel pad (Parker Laboratories Inc., Fairfield, USA). We measure the 64 center channels of a 128 element linear array transducer (L7-4, Advanced Technology Laboratories Inc., Bothell, USA) with a center frequency of 5\,MHz, a pitch of 0.3\,mm and a fractional bandwidth of 80\,\%. The number of acquisition channels was limited by our 64 channel US data acquisition system (V-1-64, Verasonics, Inc., Kirkland, USA). For this study, we used the full tuning range of our OPO and acquired PA measurements for 16 equidistant wavelengths from 680\,nm to 980\,nm in 20\,nm steps, each for four illumination positions. After firing one pulse of one wavelength in each fiber bundle the wavelength is tuned to the next in sequence. Using this 4 $\times$ 16 sequence, each MI and multispectral stack of PA images takes 640\,ms to acquire. We generally recorded the raw data for 30 such stacks for each scan. Live beamforming and visualisation with 25\,fps was performed using custom matlab scripts but this live visualisation was solely used for probe positioning and quality control (e.g.\ avoiding air inclusions under the gel pad).

\subsection{Image processing}
The acoustic reconstruction of PA images for further analysis was performed using the PA image processing module from the Medical Imaging Interaction Toolkit (MITK) \cite{kirchner2019open}. The raw data was beamformed using a delay and sum (DAS) algorithm, with a fixed speed of sound of 1480\,ms$^{-1}$ and a Hann apodization over an angle of $\pm30$ degrees. For noise reduction, the beamformed data was bandpassed. A B-Mode image was formed by using an envelope detection filter and downsampling the result to a 0.15\,mm isometric resolution. The full image processing pipeline including all relevant parameters is part of the open source appendix. The B-mode images were corrected for the mean laser pulse energy at a specific wavelength. This mean laser pulse energy correction was determined directly at the fiber bundles output before the experiments -- averaging the pulse energy for 30 laser pulses of each wavelength. For a single wavelength the variation of pulse energy was less than 3\,\%, to reduce this noise component's influence, we also averaged our PA measurements over 30 full stacks of measurements.

\subsection{Phantoms}
The phantoms used, consisted of arrays of polythene tubing (Smiths Medical International Ltd., Kent, UK) with 0.58\,mm inner diameter and 0.96\,mm outer diameter. These tubes were filled with a \emph{relative copper sulfate model} solution (as detailed in section\,2.3.1) and arranged as shown in section\,2.3.3. The relative copper (rCu) in this model is mimicking blood oxygenation (sO$_2$). 

For all the phantom experiments the background scattering medium was a fat emulsion (SMOF\-lipid 20\,\%, Fresenius Kabi, Switzerland) diluted to 1.5\,\% fat content.
To avoid errors introduced by inter-batch variations in the scattering properties of stock fat emulsions like intralipid or SMOFlipid, the optical properties of the used stock emulsion was assessed with a Time Correlated Single Photon Counting (TCSPC) technique as detailed in section\,2.3.2.

\subsubsection{Relative copper sulfate model}
\begin{figure}[hbt]
\centering
\includegraphics{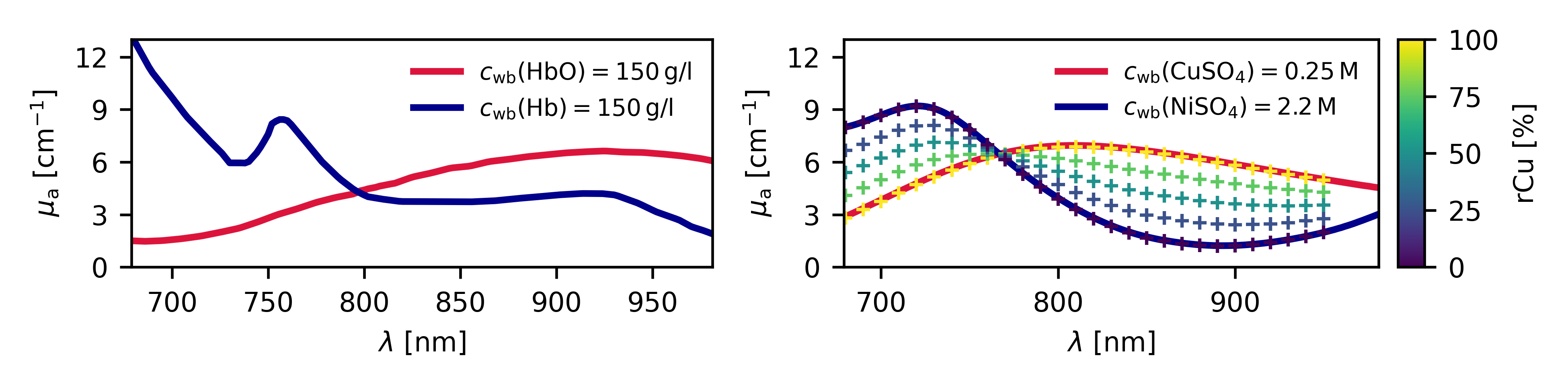}
\caption[Absorption coefficient $\mu_\textrm{a}$ spectra.]{\label{fig:refSpectra} Absorption coefficient $\mu_\textrm{a}$ spectra. Left: Oxy- and deoxyhemoglobin at whole blood concentrations $c_\textrm{wb}$(HbT) = 150\,gl$^{-1}$. Right: Copper and nickel sulfate in aqueous solution in whole blood equivalent solutions using a relative copper sulfate (rCu) model. The reference measurements of the five rCu mixtures used in our phantoms are plotted as “+”. Sulfate spectra were measured with a spectrophotometer.}
\end{figure}

The relative copper sulfate model solution was based on a 2.2 molar nickel sulfate (NiSO$_4$) water solution, produced using nickel(II) sulfate hexahydrate ($>$98\,\%, Sigma-Aldrich) and on a 0.25 molar copper sulfate (CuSO$_4$) water solution, produced using copper(II) sulfate pentahydrate ($>$98\,\%, Sigma-Aldrich) \cite{fonseca2017sulfates}. As illustrated in figure\,\ref{fig:refSpectra}, these chromophores are mimicking the NIR absorption spectra of oxy- and deoxyhemoglobin in average whole blood with a hemoglobin concentration $c_\textrm{wb}$(HbT) = 150\,gl$^{-1}$ \cite{prahl1998tabulated}. Copper and nickel sulfate were also chosen for their temporal stability and resistance to bleaching. 

The spectra of the sulfate solutions absorption coefficients $\mu_\textrm{a}$ in whole blood mimicking concentrations are defined as $c_\textrm{wb}$ (NiSO$_4$) $\coloneqq$  2.2\,M and $c_\textrm{wb}$ (CuSO$_4$) $\coloneqq$ 0.25\,M. These solutions were measured using a 2\,mm quartz cuvette (QS Hellma, Müllheim, Germany) in a UV-VIS-NIR spectrophotometer (Perkin Elmer Lambda 750, Waltham, USA), in the range of 680\,nm to 980\,nm. The scattering in this wavelength range is negligible \cite{fonseca2017sulfates}. The initial reference measurements were done in 2\,nm steps, with a 10\,s integration time and using a photomultiplier tube (PMT) sensor. 
The absorption spectroscopy measurements were repeated on the solutions after 70 days to verify their stability over time. Whenever new batches of the sulfate solutions were produced, their absorption spectra were checked against the spectra of the first batch. The solutions were corrected when they deviated from the reference spectra by more than 1\,\%.

The relative copper (rCu) in this model aims to mimic blood oxygenation (sO$_2$) and is therefore similarly defined as
\begin{equation}
\mathrm{rCu} = \frac{c_\textrm{r}(\mathrm{CuSO_4})}{c_\textrm{r}(\mathrm{CuSO_4}) + c_\textrm{r}(\mathrm{NiSO_4})},
\end{equation}
with the respective concentrations of the sulfate solutions relative to their blood mimicking base solutions
\begin{equation}
c_\textrm{r}(\mathrm{CuSO_4}) = \frac{c(\mathrm{CuSO_4})}{c_\textrm{wb}(\mathrm{CuSO_4})}\quad\text{and}\quad
c_\textrm{r}(\mathrm{NiSO_4}) = \frac{c(\mathrm{NiSO_4})}{c_\textrm{wb}(\mathrm{NiSO_4})}.
\end{equation}
For comparison, the definition of blood oxygen saturation is
\begin{equation}
\mathrm{sO}_2 = \frac{c(\mathrm{HbO}_2)}{c(\mathrm{HbO}_2) + c(\mathrm{Hb})}.
\end{equation}
While of course not following hemoglobin spectra exactly, this sulfate model is a good qualitative fit to hemoglobin and is much easier to accurately control and reproduce than the saturation of oxygen in hemoglobin. It is highly stable over time; i.e.\ over 70 days only changes smaller than 1\,\% in absorption were observed. Mimicking the blood volume fraction (bvf) in tissue we define a sulfate volume fraction (svf) in our model as $\textrm{svf} = c_\textrm{r}(\mathrm{CuSO_4}) + c_\textrm{r}(\mathrm{NiSO_4})$. The svf within the blood vessel mimicking tubing was always 100\,\%, mimicking whole blood, whereas the svf in the background was varied as detailed in section\,2.3.3.

\subsubsection{Optical property reference measurements of phantoms}
In the background medium, scattering comparable to tissue (i.e.\ $\mu_\textrm{s}'$ = 15\,cm$^{-1}$ at 750\,nm) was obtained by using a 1.5\,\% fat emulsion (diluted from SMOFlipid 20\,\%, Fresenius Kabi, Switzerland).

To ensure a reproducible and tissue mimicking scattering, the background medium was analysed with Time Correlated Single Photon Counting (TCSPC) spectroscopy. The TCSPC instrument used for the spectral analysis of the emulsions optical properties consisted of a white light supercontinuum laser (SuperK Extreme, NKT Photonics, Birkerød, Denmark) with $\approx$ 100\,ps pulse duration (varying with wavelength), running at 39\,MHz with less than 4\,mW laser output. This white light was filtered by a tunable filter (SuperK Varia, NKT Photonics, Birkerød, Denmark), which was tuned in a range from 600\,nm to 840\,nm in 20\,nm steps, with a bandwidth of 10\,nm; 840\,nm being the maximum of the tunable filter’s range.
A single-photon avalanche diode (MDP PDM Series, Micro Photon Devices, Bolzano, Italy) was used to detect single photons. The diode has a prolonged dead time of $\approx$ 80\,ns after a photon detection. Because of that, the photon detection rate was kept sufficiently low to make photon detection events during the dead time unlikely. We ensured a detection rate lower than $10^5$\,s$^{-1}$ ($\ll 1/80$\,ns), making a correction for missed photons during the dead time unnecessary. The distributions of times of flight (DTOFs) were recorded with single photon counting electronics (SPC-160, Becker \& Hickl GmbH, Berlin, Germany). Source and detector fiber were fixed in blunted hypodermic needles for stability. The laser pulse shape, temporal dispersion in the optical fibers, as well as the response of the detector were characterized in the overall instrument response function (irf), yielding a FWHM of $\approx$ 140\,ps overall, varying with wavelength.
The source and detection fibers were placed perpendicular to the surface of the sample medium and immersed in the medium by 0.5\,mm. To reduce the detection of early arriving photons a carbon fiber mesh blocker was placed into the direct path, at a distance of 6\,mm from the source fiber (dimension: 1\,mm depth, 4\,mm width, 0.4\,mm thickness). We measured the SMOFlipid 1.5\,\% medium in an 8\,cm radius, 10\,cm deep beaker, with the fibers at the center. This is a sufficiently large volume to be approximated as a semi-infinite medium for the analytic diffusion model. The resulting media were both measured with a source detector separation $\rho$ = 20\,mm, for each wavelength until at least $10^7$ photons were detected. For some wavelengths, the laser needed to be attenuated in order to keep the photon detection rate below $10^5$\,s$^{-1}$. This acquisition protocol ensured a high signal-to-noise ratio and allowed us to fit our diffusion model only to late arriving photons where the diffusion approximation is more accurate. For the phantom experiments two bottles of a new batch of SMOFlipid were used -- both batches and bottles were measured independently prior to experiments to avoid hidden variations in the background medium.

\begin{figure}[hbt]
\centering
\includegraphics{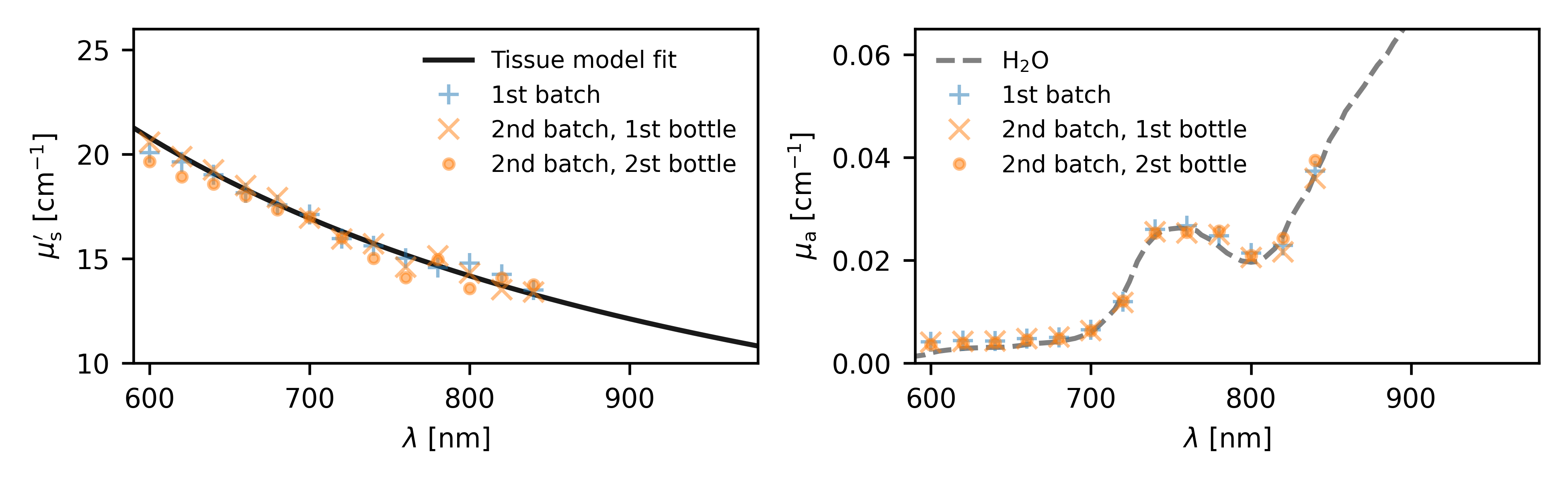}
\caption[Optical properties of the uncolored phantom background medium.]{\label{fig:scattering} Optical properties of the uncolored phantom background medium. Single data points are diffusion model results from measurements with the TCSPC spectroscopy instrument of a 1.5\,\% fat emulsion (diluted from SMOFlipid 20\,\%). Validation Phantoms cf. figure\,\ref{fig:phantoms}a were constructed with the “2nd batch, 1st bottle”. A generic tissue scattering model (eq.\,\ref{eq:muspModel}) fitted to this first bottle measurement was used to set the background scattering properties for the Monte Carlo simulations. The “2nd bottle” was used for the background media in the test phantoms. The absorption results are shown together with water absorption\cite{segelstein1981complex}.}
\end{figure}

An analytic diffusion model\cite{groenhuis1983scattering} with an extrapolated boundary condition for a semi-infinite medium \cite{cubeddu1999compact, hielscher1995influence} was convolved with the corresponding irf for each wavelength $\lambda$. The results were then fitted to the measured histograms of the single photon arrival times, yielding a series of tuples $(\mu'_\textrm{s}{}^\textrm{SPC}(\lambda),$ $ \mu_\textrm{a}{}^\textrm{SPC}(\lambda))$. Our tunable filter was limited in range to a maximum wavelength 840\,nm but we needed credible $\mu'_\textrm{s}$ values up to 980\,nm for the optical forward simulations. Therefore, a generic tissue model (eq.\,\ref{eq:muspModel}) from the mcxyz framework\cite{jacques2014coupling} was used to expand and define the scattering properties within the optical forward simulation.
\begin{equation}
\label{eq:muspModel}
\mu'_\textrm{s}(\lambda) = \mu'_\textrm{s}{}_{500}\cdot(f_\textrm{ray}\cdot(\lambda/500\,\textrm{nm})^{-4} + (1-f_\textrm{ray})\cdot(\lambda/500\,\textrm{nm})^{-b_\textrm{mie}})
\end{equation}

with $\mu'_\textrm{s}{}_{500} = 42.4$\,cm$^{-1}$ the initial guess for $\mu'_\textrm{s}$ at 500\,nm, $f_\textrm{ray}$ = 0.62 the initial guess for fraction of Rayleigh scattering at 500\,nm, $b_\textrm{mie}$  = 1.0  the initial guess for the scatter power for Mie scattering. This was fitted to the TCSPC data with a least squares fit -- the entire data processing pipeline with all parameters is part of the open source code supplement. The resulting fits are shown in figure\,\ref{fig:scattering}.

\subsubsection{Phantom data sets}
\begin{SCfigure}[50][hbt]
\includegraphics{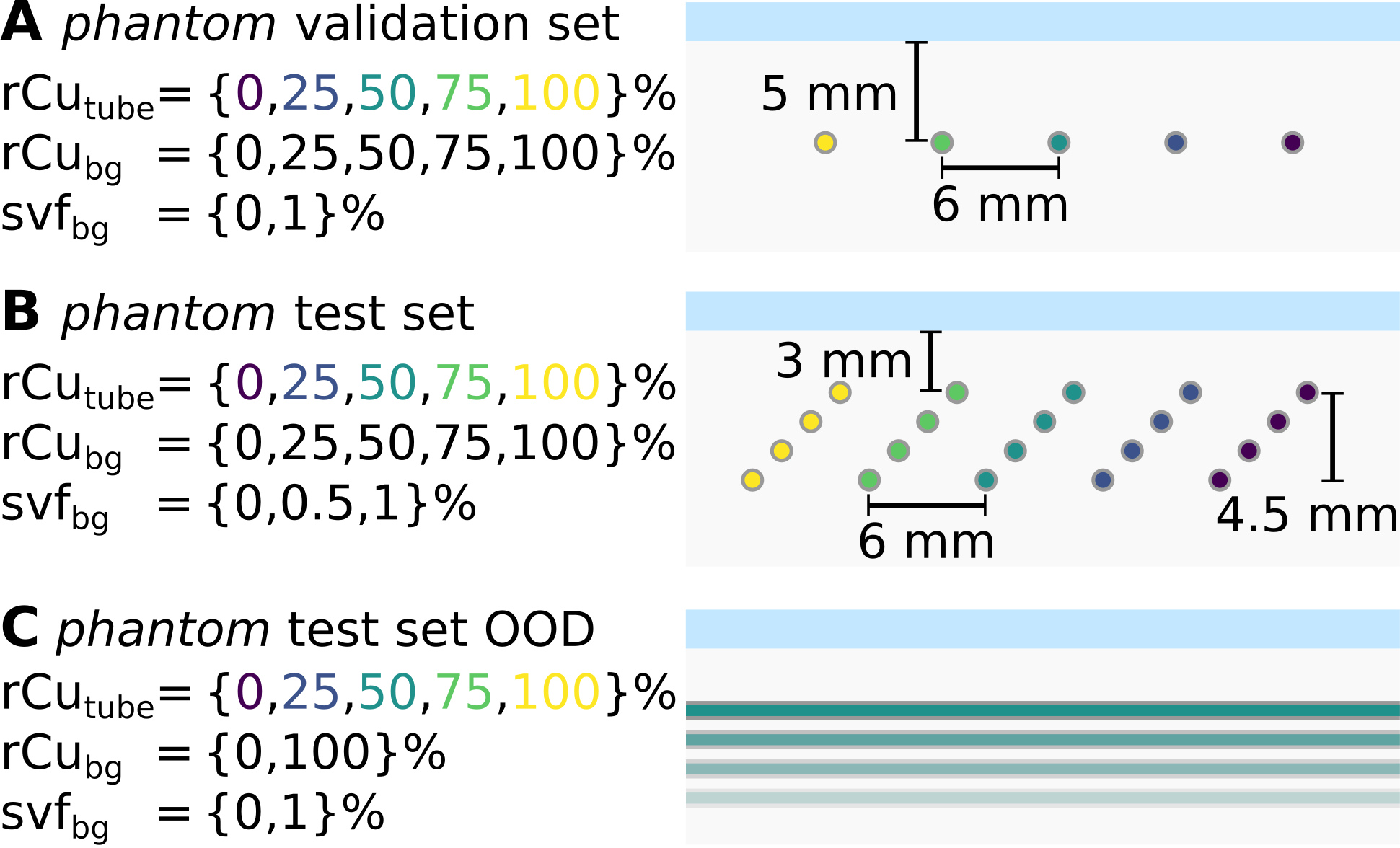}
\caption[Cross sections of the phantom sets.]{\label{fig:phantoms} Cross sections of the phantom data sets, with denoted parameters: relative copper sulfate model in the tubes rCu$_\textrm{tube}$, in the background medium rCu$_\textrm{bg}$ and sulfate model volume fraction in the background medium svf$_\textrm{bg}$. \textbf{A} The validation phantoms with five single tubes. \textbf{B} The main test phantoms. \textbf{C} The test phantoms in longitudinal scan direction and thereby somewhat out-of-distribution (OOD) of the training data. The shown 2D cross sections correspond to the imaging plane. In set A and B the tubes run perpendicular to the imaging plane. Phantom test C has the same geometry as set B, with the imaging plane rotated by 90 degrees to yield longitudinal scans instead of transversal scans of the tubes.}
\end{SCfigure}

Three sets of phantoms (A,B,C) were produced, with different layout as shown in figure\,\ref{fig:phantoms}. All phantoms use polythene tubing filled with the \emph{relative copper sulfate model} solution as target structures. The phantom backgrounds consist of a 1.5\,\% fat emulsion with added sulfates.

Phantom layout A was measured as a validation data set for hyperparameter tuning of the machine learning models and validation of image reconstruction as well as parameter tuning in the Monte Carlo simulations. Layouts B and C were exclusively measured as test data sets. Phantom test set B are expected to be within the distribution of the simulation parameters (cf.\,figure\,\ref{fig:inSilico}). Phantom test set C however consists only of longitudinal scans w.r.t.\ the tube orientation. Because the orientation of the illumination positions changes with the imaging plane, set C was illuminated along the tubing. The measurements in set C are therefore expected to be out-of-distribution (OOD) with respect to the Monte Carlo simulated training sets. As detailed in the next section, simulations were exclusively performed for transversal orientation of the tubing.

\noindent The phantom data sets contain 164 multispectral MI PA scans from 115 scan configurations as follows:
\begin{enumerate}[label=\textbf{\Alph*}]
\item 30 scan configurations as laid out in figure\,\ref{fig:phantoms}A: Six phantom configurations: one with only a 1.5\,\% SMOFlipid background solution, five with an added 1\,\% sulfate volume fraction (svf) background with relative copper rCu$_\textrm{bg}$ set to \{0, 25, 50, 75, 100\}\,\%. On each of these six configurations five MI multispectral scans were performed centering the transducer on each of the tubes with rCu$_\textrm{tube}$ = \{0, 25, 50, 75, 100\}\,\%.
\item 55 scan configurations as laid out in figure\,\ref{fig:phantoms}B: Eleven phantom configurations: one with only a 1.5\,\% SMOFlipid background solution, five with an added 1\,\% svf background with rCu$_\textrm{bg}$ = \{0, 25, 50, 75, 100\}\,\% and five with a 0.5\,\% svf. On these eleven phantom configurations MI multispectral scans were performed centering on each of the five four-tube-arrays with rCu$_\textrm{tube}$ = \{0, 25, 50, 75, 100\}\,\%. For each four-tube-array, two regions of interest (ROI) (one containing the two lower and one the two upper tubes) were analysed separately. The imaging plane was positioned for transversal scans of the tubes.
\item 30 scan configurations as laid out in figure\,\ref{fig:phantoms}C: Three phantom configurations: one with only a 1.5\,\% SMOFlipid background solution, two with an added 1\,\% svf background with rCu$_\textrm{bg}$ = \{0, 100\}\,\%. On these three phantom configurations MI multispectral scans were performed with each of the five shallowest tubes, and each of the five deepest tubes of the four-tube-arrays in the imaging plane, with rCu$_\textrm{tube}$ = \{0, 25, 50, 75, 100\}\,\%. The imaging plane was positioned for longitudinal scans of the tubes.
\end{enumerate}

All scan configurations were scanned for 19.2\,s yielding 30 MI and multispectral sequences. Due to the limited field of view of our US system (parallel read-out of 64 channels on a 19.2\,mm linear array) we repositioned the probe between acquisitions -- i.e.\ measuring five scan positions for phantom geometries A and B. The center of the linear transducer was always placed above the center of the targeted tubes. Scans with technical difficulties such as frame drops or wrong positioning were discarded in post processing, this affected one of the 115 scan configurations: the rCu$_\textrm{tube}$ = 100\,\%, rCu$_\textrm{bg}$ = 50\,\%, svf = 0.5\,\% was discarded for erroneous positioning. All scans of Phantom geometry C were performed twice. The svf = 0 scans on phantom geometry B were performed five times on different days as a baseline measurement. The total phantom data set consists of 164 scans.

It is important to note that both copper and nickel sulfate act as a demulsifier when mixed with the diluted SMOFlipid background, or any other fat in water emulsion. Phases will form and the bulk optical properties will change significantly within tens of seconds. To avoid the forming of phases, the background medium with added sulfates was continuously stirred with a magnetic stirrer during all the measurements.

\subsection{Optical forward simulations}
\begin{SCfigure}[50][hbt]
\centering
\begin{tabular}{c}
\includegraphics{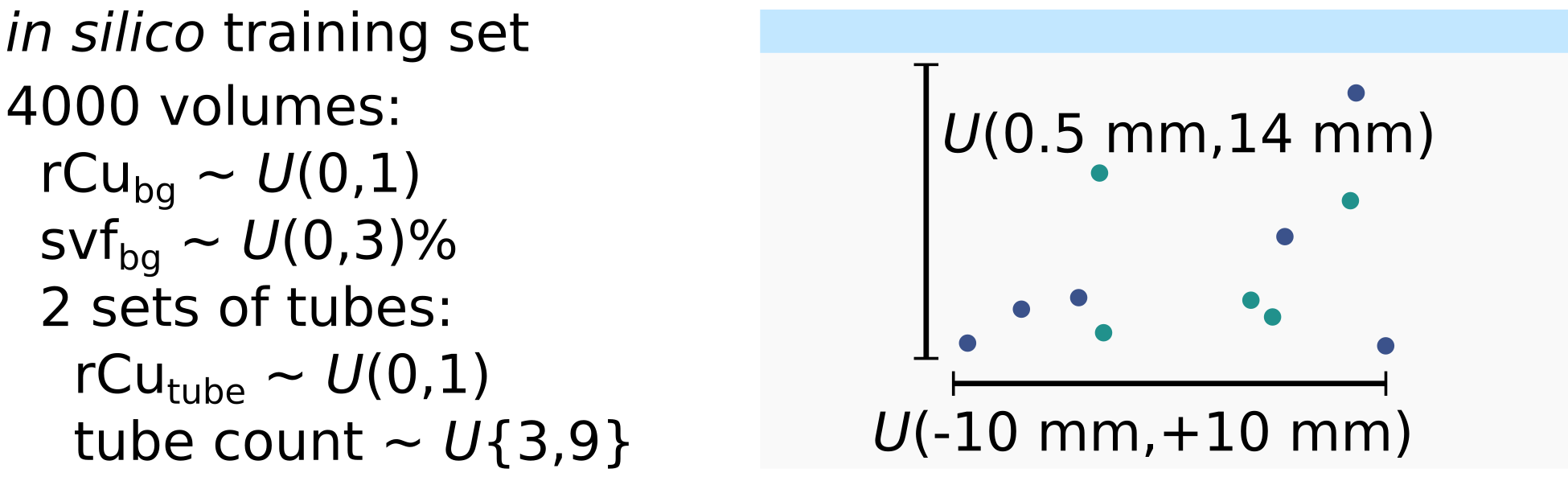}
\end{tabular}
\caption[Cross sections of the phantom geometries.]{\label{fig:inSilico} The \emph{in silico} training data set consists of 4000 volumes, simulated with Monte Carlo simulations, modeling the geometry of the real multiple illumination (MI) setup. An additional 1000 volume test set is kept separate. Each volume has two sets of tubes each with a random number of tubes, uniformly distributed as specified. Tube and background relative copper (rCu) as well as background sulfate volume fraction (svf) are also drawn from uniform random distributions $U$.}
\end{SCfigure}

As an optical forward model we used GPU accelerated Monte Carlo simulations to generate ground truth  multispectral stacks of the absorbed energy distributions $H(\boldsymbol{x}, \boldsymbol{\lambda})$. Figure\,\ref{fig:inSilico} illustrates the layout of the Monte Carlo simulated volumes. The simulated data set consists of a 4000 volume training set and a separate 1000 volume test set. For each volume 16 wavelength and four positions of illumination were simulated, modeled on the real MI PA imaging sequences. The simulations were performed with the open source mcx toolkit \cite{fang2009monte} and we used the ippai framework for the illumination modelling and data organisation. In all data sets, each volume has two sets of tubes with the tube count drawn from a discrete uniform distribution $U\{3,9\}$, uniformly distributed in the volume as specified in figure\,\ref{fig:inSilico}. Tube and background rCu are drawn from a continuous uniform random distribution $U(0,1)$. All tubes were set to a radius of 0.4\,mm.
The wavelength dependent background scattering parameters were set to the tissue model results from the fit of the TCSPC measurements to equation\,\ref{eq:muspModel}. The background svf was drawn from $U(0, 3)\%$. Each simulation was performed with $10^8$ launched photon packets. Running these simulations on a high performance computing cluster we used mostly 1080 GTX  (Nvidia, Santa Clara, USA) GPUs, with which a single wavelength, single illumination position, simulation took approximately two minutes. All simulations for the test and training sets used a combined 400 days of GPU time. This was made feasible by usually running 40 GPUs in parallel.
It should be noted that this seemingly excessive simulation time was chosen after simulation results with $10^7$ photon packets proved too noisy. This was evaluated prior to the presented \emph{in silico} data sets. Initial hyperparameter tuning was also performed on two \emph{in silico} data sets, simulated with $10^7$ photon packets. These data sets are part of the supplemental data.

\subsection{Machine learning algorithms}
The estimation of a sO$_2$ or rCu value from a measured spectrum is a regression problem. The usual approach to this problem in PA imaging is linear spectral unmixing (LU)\cite{tzoumas_spectral_2017, kirchner2019open}. For one pixel, the PA signal spectrum $\boldsymbol{S}(\boldsymbol{\lambda})$ is measured at a set of wavelengths $\boldsymbol{\lambda}$. This sampled PA signal spectrum $\boldsymbol{S}$ is then fitted to a linear combination of known absorption spectra. Here, LU is performed numerically using an iterative least squares solver implemented in Python's scipy.optimize submodule. These LU estimations (rCu$_\textrm{est}^\textrm{LU}$) are given throughout the results section as a reference. 

We also compare our results to LSD, a type of machine learning algorithm. LSD also aims to estimate sO$_2$ or rCu from the the same single illumination PA signal spectra $\boldsymbol{S}$ measured at wavelengths $\boldsymbol{\lambda}$. Similar to prior implementations, our modified LSD models are machine learning algorithms that are trained on large amounts of simulated absorbed energy spectra labeled with ground truth rCu. Before training, each absorbed energy spectrum is normalized with its L1 norm to $\hat{\boldsymbol{H}}(\boldsymbol{\lambda})$. This normalization makes them equivalent to a normalized PA signal spectrum $\hat{\boldsymbol{S}}(\boldsymbol{\lambda})$. This is because we can assume that for a signal spectrum $\boldsymbol{S}$ at a position $\boldsymbol{x}_0$
\begin{align}
    \boldsymbol{S}(\boldsymbol{x}_0, \boldsymbol{\lambda}) &= \Gamma(\boldsymbol{x_0}) \cdot A(\boldsymbol{x}_0) \cdot \boldsymbol{H}(\boldsymbol{x}_0, \boldsymbol{\lambda}) \\
    \Rightarrow \hat{\boldsymbol{S}}(\boldsymbol{x}_0, \boldsymbol{\lambda}) &\approx \hat{\boldsymbol{H}}(\boldsymbol{x}_0, \boldsymbol{\lambda})\textrm{.}
\end{align}

Assuming a linear acoustic reconstruction like delay and sum, $A(\boldsymbol{x}_0)$ is a spatially varying but wavelength independent factor introduced by the imperfect acoustic reconstruction, the instrument response and the calibration. $\Gamma(\boldsymbol{x_0})$, as a material property is also independent of the illumination wavelength\cite{fonseca2017sulfates}. 
The LSD model, which was trained on the \emph{in silico} training set tuples $(\hat{\boldsymbol{H}}, \textrm{rCu}_\textrm{tube})$ is then presented (1) unseen \emph{in silico} test set spectra  $\hat{\boldsymbol{H}}$ or (2) spectra $\hat{\boldsymbol{S}}$ from an unseen phantom data test set to estimate the corresponding rCu$_\textrm{est}^\textrm{LSD}$. 

Note that $A$ actually does depend on the fluence distribution $\boldsymbol{\phi}(\boldsymbol{x}, \mu_\textrm{a}(\boldsymbol{x}, \boldsymbol{\lambda}), \mu_\textrm{s}'(\boldsymbol{x}, \boldsymbol{\lambda}))$. A varying optical wavelength may lead to different acoustic spectra of the PA signal corresponding to the same structure, due to different spatial distributions in the absorbed energy. Our assumption is that this effect is small compared to the spectral coloring introduced directly by the fluence term in
\begin{align}
    \boldsymbol{H}(\boldsymbol{x}_0, \boldsymbol{\lambda}) &= \boldsymbol{\phi}\big(\boldsymbol{x}_0, \mu_\textrm{a}(\boldsymbol{x}, \boldsymbol{\lambda}), \mu_\textrm{s}'(\boldsymbol{x}, \boldsymbol{\lambda})\big) \cdot \boldsymbol{\mu}_\textrm{a}(\boldsymbol{x}_0, \boldsymbol{\lambda}).
\end{align}

\begin{figure}[hbt]
\centering
\includegraphics{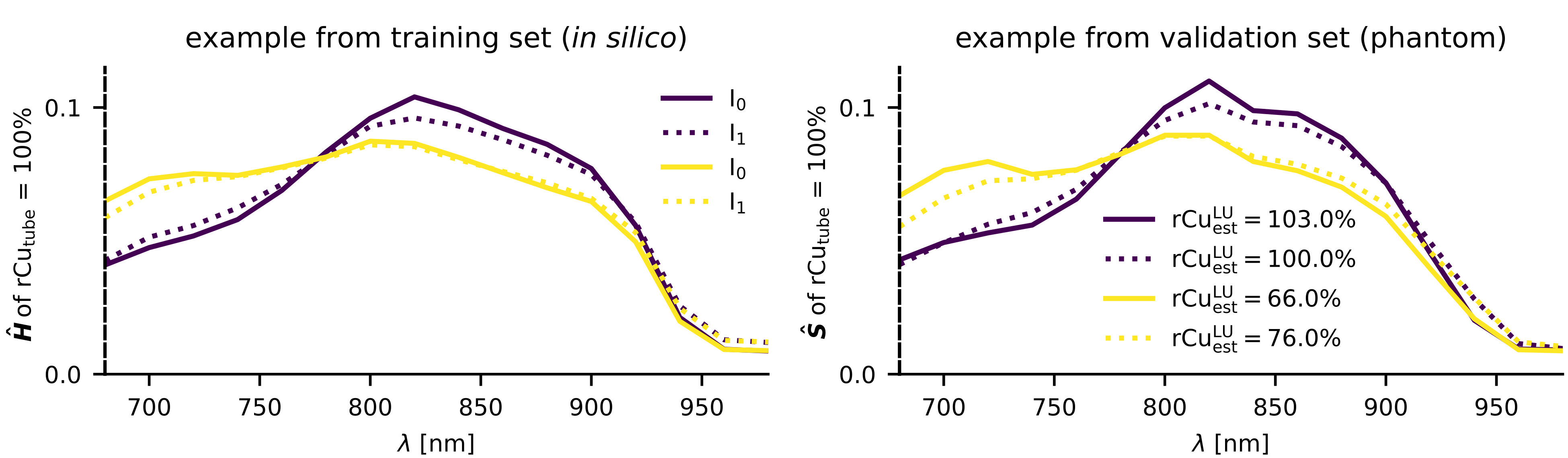}
\caption[Examples of L1 normalized spectra.]{\label{fig:miSpectraIllustration}Examples for spectral coloring in L1 normalized spectra of an absorber with rCu$_\textrm{tube}$ = 100\,\%. Comparison between (left) \emph{in silico} absorbed energy spectra $\hat{\boldsymbol{H}}$ and (right) phantom PA signal spectra $\hat{\boldsymbol{S}}$. Spectra for two background media are shown: rCu$_\textrm{bg}$ = 100\,\% (yellow), rCu$_\textrm{bg}$ = 0\,\% (dark); with svf = 1\,\%. The spectra for two illumination positions $I_0$ (line) and $I_1$ (dotted) are shown as an example for two of the four illuminations. Systematic changes in the spectral coloring can be seen for different background media and illumination positions. These changes are qualitatively similar for $\hat{\boldsymbol{H}}$ and $\hat{\boldsymbol{S}}$. On the validation phantom examples LU estimations for single spectra rCu$^\textrm{LU}_\textrm{est}$ are listed -- spectral coloring can cause large estimation errors relative to the rCu$_\textrm{tube}$ = 100\,\% ground truth.}
\end{figure}

For MI-LSD we have multiple such normalized spectra $\hat{\boldsymbol{S}}$ as input variables. For illustration, figure\,\ref{fig:miSpectraIllustration} shows spectra of the same pixel in an absorber with rCu$_\textrm{tube}$ = 100\,\% with two example illuminations $I_0$, $I_1$ and for two backgrounds with rCu$_\textrm{bg} =$ $\{$0\,\%,\,100\,\%$\}$ and svf = 1\,\%. The difference in background absorption causes a different spectral coloring but so does a variation of the illumination position. We hypothesise that training our machine learning algorithms on i.e.\ four such spectra will allow us a more accurate and/or more robust estimation compared to LU and LSD.

Two types of machine learning algorithms were employed for spectral decoloring: feed forward neural networks (NN) and random forests (RF). Training of the MI-LSD models includes mirrored illumination positions for each volume as a minor data augmentation. Sorting the training data illumination position spectra stacks by their L1 norm before training was also investigated but did not prove beneficial on the validation data.

\subsubsection{Feed forward neural networks}
Feed forward neural networks were previously used for LSD implementations \cite{grohl2021learned, grohl2019estimation}. We used this state-of-the-art NN architecture as a starting point and further tuned the hyperparameters on the training and validation sets. Doing so we mainly found the dropout layers of previous implementations to be counterproductive -- dropout leading to a much lower precision on the validation set. The two final NNs used for both LSD and MI-LSD consisted of four hidden layers with twice the size of the input layer (16 for LSD and 64 for MI-LSD), all with leaky ReLu activation layers (and for comparison, dropout layers). For comparison to the previous implementation, additional results for a dropout in the dropout layers with probability $p = 0.2$ are presented in the supplemental figures 35-50 and 66-72. In the main results no dropout was used ($p = 0$). We segmented all vessels in the 4000 volume training set and trained on the segmented 1,052,152 simulated MI signal spectra for 100 epochs. As in the previous implementation we used a batch size of $10^5$ and a learning rate of $10^{-2} \cdot 0.9^{\mathtt{epoch}/2}$. We trained the algorithms on a RTX 2060 Super GPU (Nvidia, Santa Clara, USA) and used the CPU for inference.

\subsubsection{Random forest regression}
We also investigated random forest regression\cite{breiman2001random}, usually a highly accurate learning algorithm for  regression problems with few dimensions\cite{kirchner_context_2018}. RFs are also usually less impacted by noise models, they especially should not overfit on noise\cite{breiman2001random}. This should prove useful as we did not try to model a realistic wavelength dependent noise. We used the python scikit-learn v0.23 implementation of random forest regressors using 100 trees with a maximum depth of 30 to limit memory consumption. Further parameters were set to default.

\section{Results}
We first show some qualitative comparisons between \emph{in silico} and phantom data and then present the performance of our trained RF and NN models on our \emph{in silico} test set and the two phantom test sets. 

The hyperparameters of the machine learning models were tuned on the phantom validation set. The models that performed best on our validation data were used to estimate rCu from the test sets. These models are presented in the results. For further information, all estimations for all models (on the validation set and for each single test measurement) can be found in the supplemental figures; representative examples are shown here.

We compare MI-LSD with LSD and LU. Comparing methods based on a single measurement with methods based on multiple measurements should generally lead to improved accuracy in the multiple measurement method simply due to an increase in signal-to-noise ratio (SNR). In order to more fairly compare MI-LSD to the \emph{single spectrum} methods like LU and LSD, we estimated LSD and LU results on the reconstructed signals, averaged over the four illuminations. By using this \emph{averaged illumination} spectrum as input for LU and LSD, we can compare methods for the same delivered energy during the same time -- giving no method an inherent SNR advantage. LSD was also trained on \emph{in silico} data using the same \emph{averaged illumination} spectra from four simulated illuminations.
\begin{figure}[hbt]
\centering
\includegraphics{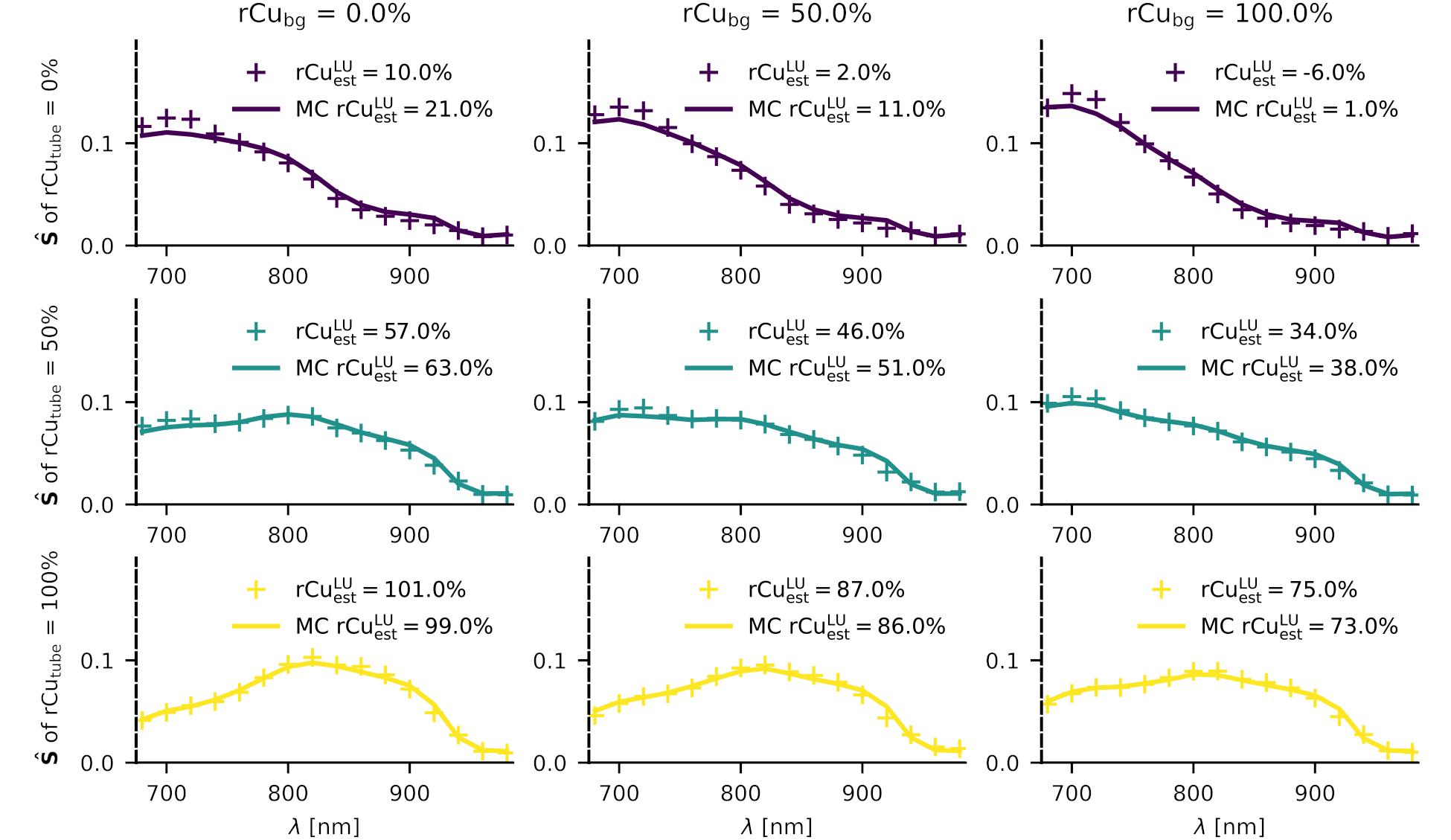}
\caption[Qualitative comparison between a validation phantom and its digital twin.]{\label{fig:initialValidation} Qualitative comparison between spectra in a validation phantom (for svf = 0) and its digital twin from Monte Carlo (MC) simulations, showing the effects of various spectral coloring on the mean illumination spectra. Relative copper rCu$_\textrm{tube}$ is varied in the target tube (up-down) and the background medium rCu$_\textrm{bg}$ (left-right). For reference, linear unmixing (LU) rCu estimates are given for each spectrum.}
\end{figure}

In addition to using the validation data set for hyperparameter tuning, we also qualitatively compared a set of our measurements to Monte Carlo simulations of one of the validation phantoms -- creating an exact \emph{in silico} representation of the light propagation in the validation phantom. Figure\,\ref{fig:initialValidation} serves as a qualitative (phantom to \emph{in silico}) comparison for some of the averaged illumination spectra. 

\begin{figure}[H]
\centering
\includegraphics{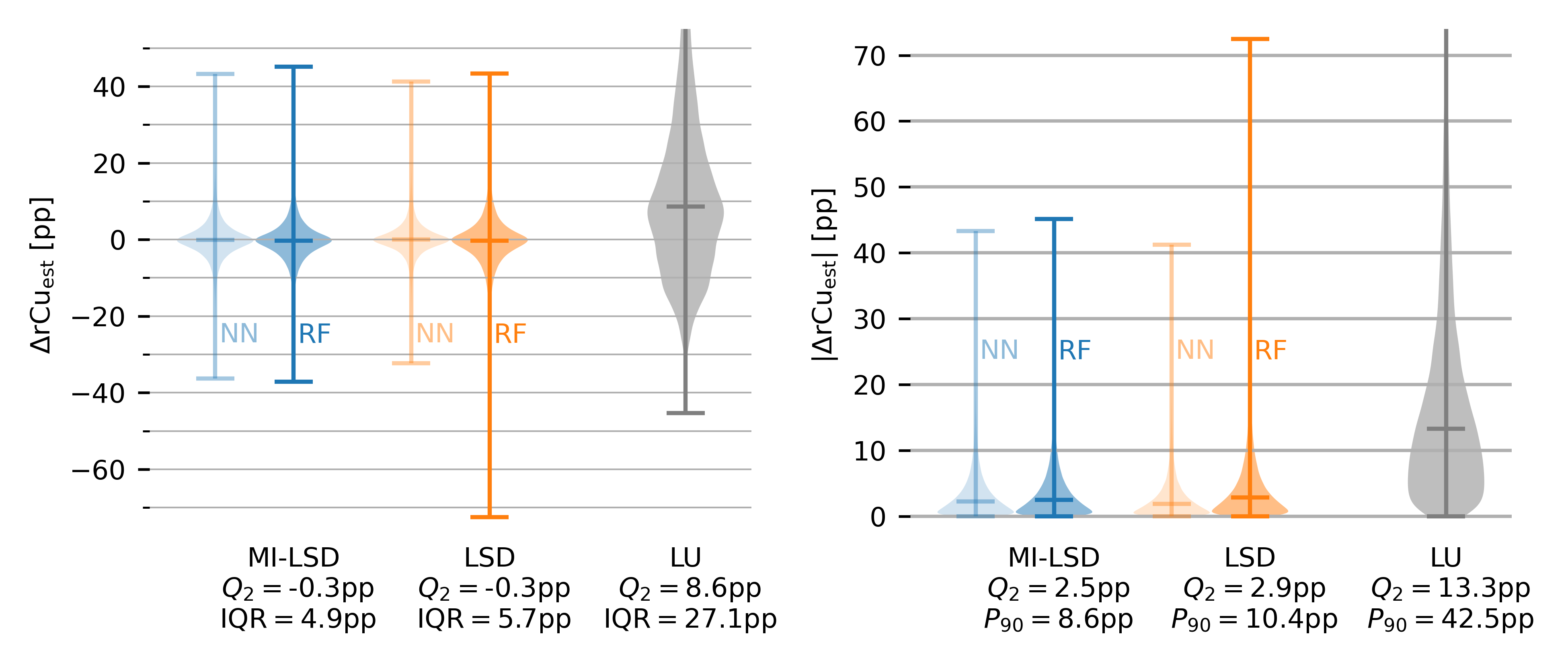}
\caption[Error distributions of the \emph{in silico} test set.]{\label{fig:errorsInSilicoTest} Error distributions of the \emph{in silico} test set cf.\ figure\,\ref{fig:inSilico}. rCu estimation errors $\Delta \textrm{rCu}_\textrm{est}$ are shown left, their absolutes right. Blue shows the rCu estimators using multiple illumination learned spectral decoloring (MI-LSD), orange the estimators using learned spectral decoloring (LSD) and gray is the linear spectral unmixing (LU) reference. Medians $Q_2$ of the error distributions are shown, together with interquartile ranges (IQR) and 90 percentiles $P_{90}$. The two feed forward neural network (NN) models and the two random forest (RF) models all have median absolute errors below 3 percentage points.}
\end{figure}

We report the estimation error distributions on the three distinct test sets. Reported are rCu estimation errors $\Delta \textrm{rCu}_\textrm{est} = \textrm{rCu}_\textrm{est} - \textrm{rCu}_\textrm{tube}$ and their absolutes $|\Delta \textrm{rCu}_\textrm{est}|$, with $\textrm{rCu}_\textrm{tube}$ being the ground truth rCu in the tube. 
In the \emph{in silico} test set all selected models are in close agreement. As shown in figure\,\ref{fig:errorsInSilicoTest}, both LSD and MI-LSD estimations with both RFs and NNs yield median $Q_2$ absolute estimation errors of less than 3 percentage points (pp). With the lower 90 percentile $P_{90}$ for MI-LSD showing fewer outliers and a potentially higher accuracy.
As expected, estimation with all used models is very fast compared to LU. Inference on CPU for all the 266,105 samples in the \emph{in silico} test set took 1.6\,s for RF MI-LSD, 1.3\,s for RF LSD, 0.2\,s for NN MI-LSD, 0.04\,s for NN LSD, compared to 642\,s for LU.

\begin{figure}[tbh]
\centering
\includegraphics{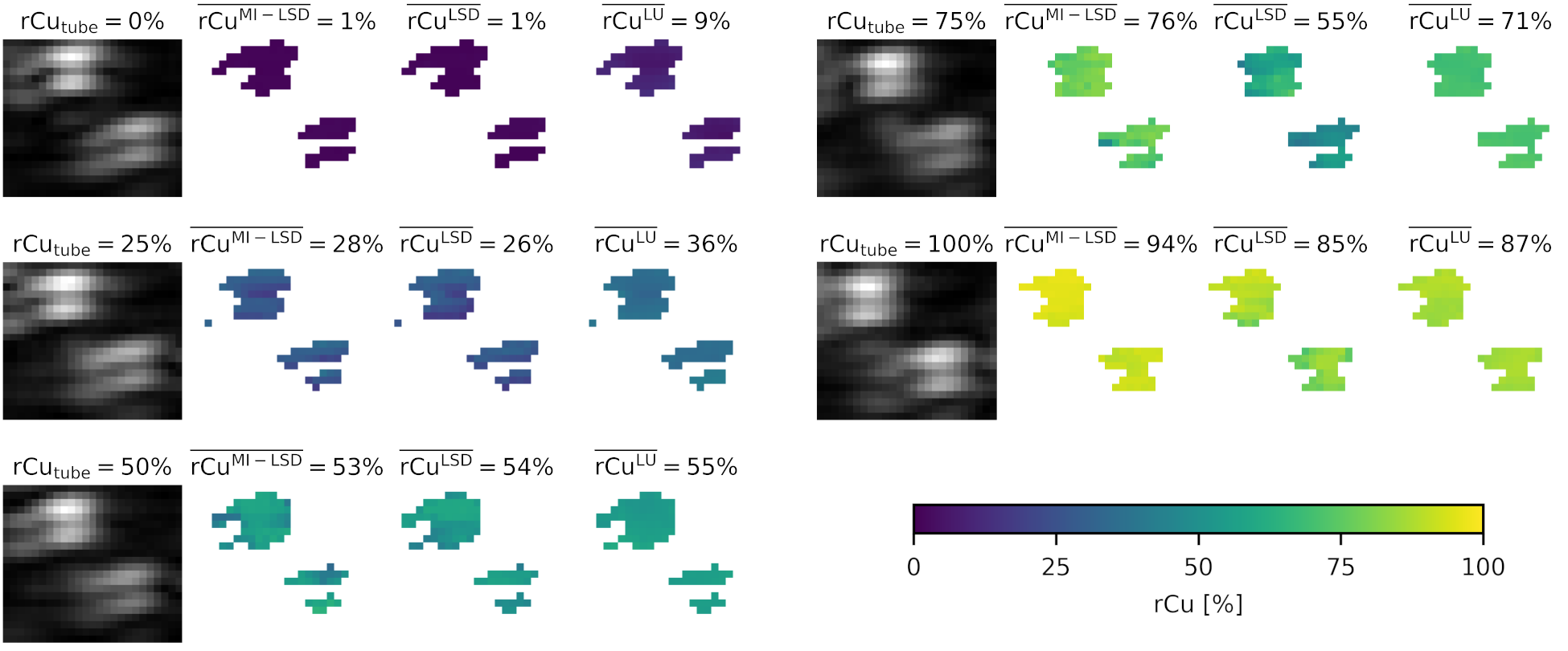}
\caption[Example region of interest (ROI) in the phantom test set B with the estimation results for MI-LSD, LSD and LU.]{\label{fig:examplesPhantomTestB_lower} Example region of interest (ROI) in the phantom test set B with the estimation results for MI-LSD, LSD and LU. Shown are the lower two tubes of a four tube phantom with svf = 0.5\,\% and rCu$_\textrm{bg}$ = 25\,\%.
Mean PA signal is shown left, with the ground truth rCu$_\textrm{tube}$ annotated. The mean rCu estimate $\overline{\textrm{rCu}}$ over the ROI is noted for the three estimators.}
\end{figure}
\begin{figure}[hbt]
\centering
\includegraphics{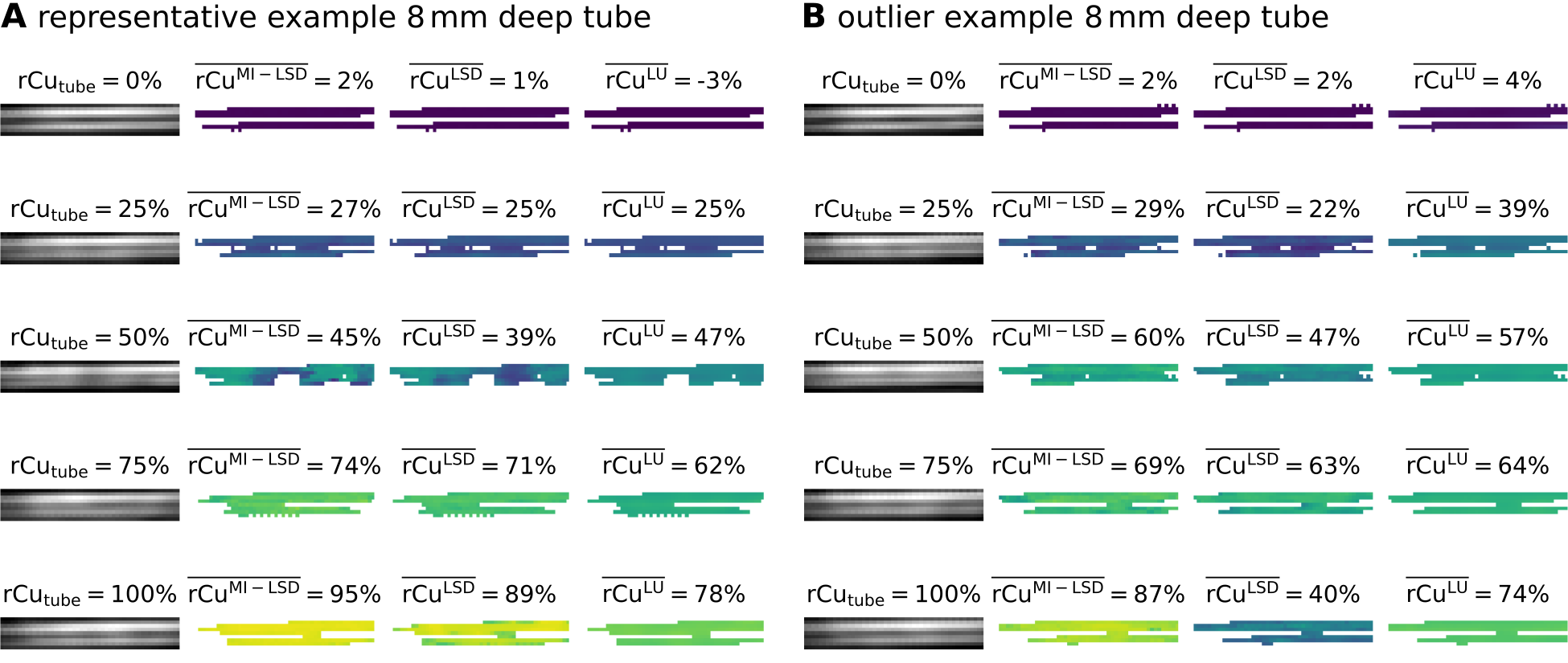}
\caption[Example regions of interest (ROI) 8\,mm deep  in the phantom test set C with estimation results for MI-LSD, LSD and LU.]{\label{fig:examplesPhantomTestC} Example regions of interest (ROI) 8\,mm deep in the phantom test set C with estimation results for MI-LSD, LSD and LU. Shown are two times five imaged tubes with their rCu$_\textrm{tube}$ annotated above their mean PA signal. The mean rCu estimate $\overline{\textrm{rCu}}$ over the ROI is noted for the three estimators.
The ROIs for two sets of phantoms are shown. \textbf{A} a representative result (rCu$_\textrm{bg}$ = 100\,\%, 1\,\% svf background), \textbf{B} a result with outlier estimation errors (0\,\% svf background). LSD has highest estimation errors  in deep vessels and in phantoms with no added sulfates in the background medium, i.e.\ in \textbf{B} for rCu$_\textrm{tube}$ = 100\,\%.}
\end{figure}

From phantom test set B, \emph{tube} signal was segmented by thresholding. In each reconstructed MI-multispectral PA image stack two ROIs were chosen, one containing the two upper tubes and one containing the lower two tubes. One such lower tubes ROI is shown in figure\,\ref{fig:examplesPhantomTestB_lower}. Each ROI has a fixed size of 3.75\,mm times 3.3\,mm. The 15\,\% highest mean (over all wavelengths and illuminations) PA signal pixels in each ROI were segmented as \emph{tube} and rCu was estimated from the MI-multispectral PA signals in all pixels of these \emph{tube} signal areas. The ROIs were thresholded separately in order to get an equal number of lower tube samples into the test set. A thresholding on the entire image or a larger ROI, using a lower cut-off percentage, would lead to more clutter and noise in the test set and the lower tubes being underrepresented in the test set.

From phantom test set C, tube signal was segmented in a similar fashion: from each reconstructed MI-multispectral PA signal image stack a ROI of fixed size (7.5\,mm times 1.5\,mm) was selected, containing either the upper tube or the lower tube. Two such lower tubes example ROIs are shown in figure\,\ref{fig:examplesPhantomTestC}A\&B for varying reference rCu$_\textrm{tube}$. Within these ROIs the 50\,\% highest mean (over all wavelengths and illuminations) PA signal pixels were segmented as \emph{tube}. rCu was then estimated from the MI-multispectral PA signals in all pixels within these \emph{tube} signal locations.

\begin{figure}[hbt]
\centering
\includegraphics{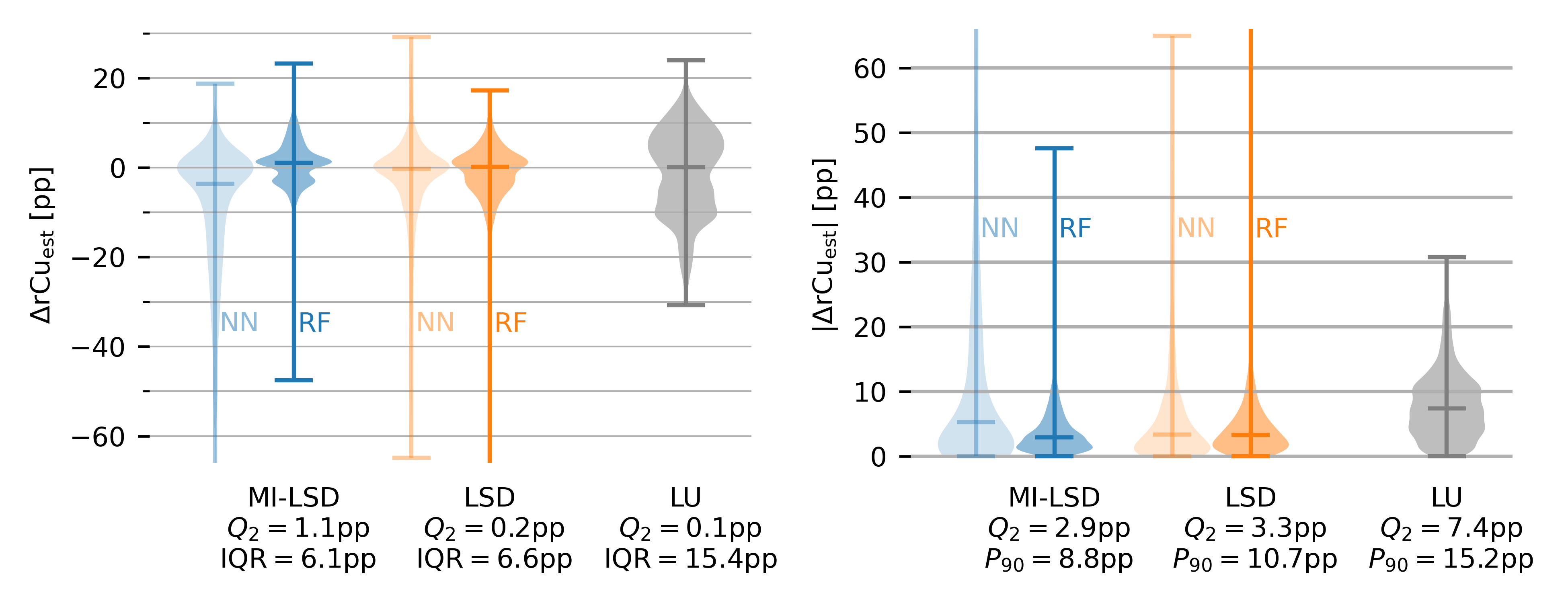}
\caption[Error distributions of the phantom test set B.]{\label{fig:errorsPhantomTestB} Error distributions of the phantom test set B (cf.\ figure\,\ref{fig:phantoms}B). rCu estimation errors $\Delta \textrm{rCu}_\textrm{est}$ are shown left, their absolutes right. Blue shows the rCu estimators using multiple illumination learned spectral decoloring (MI-LSD), orange the estimators using learned spectral decoloring (LSD) and gray is the linear spectral unmixing (LU) reference. Medians $Q_2$ of the error distributions are shown, together with interquartile ranges (IQR) and 90 percentiles $P_{90}$. The feed forward neural network (NN) models performed similar to the random forest (RF) models for the LSD method but underperformed for MI-LSD.}
\end{figure}

\begin{figure}[hbt]
\centering
\includegraphics{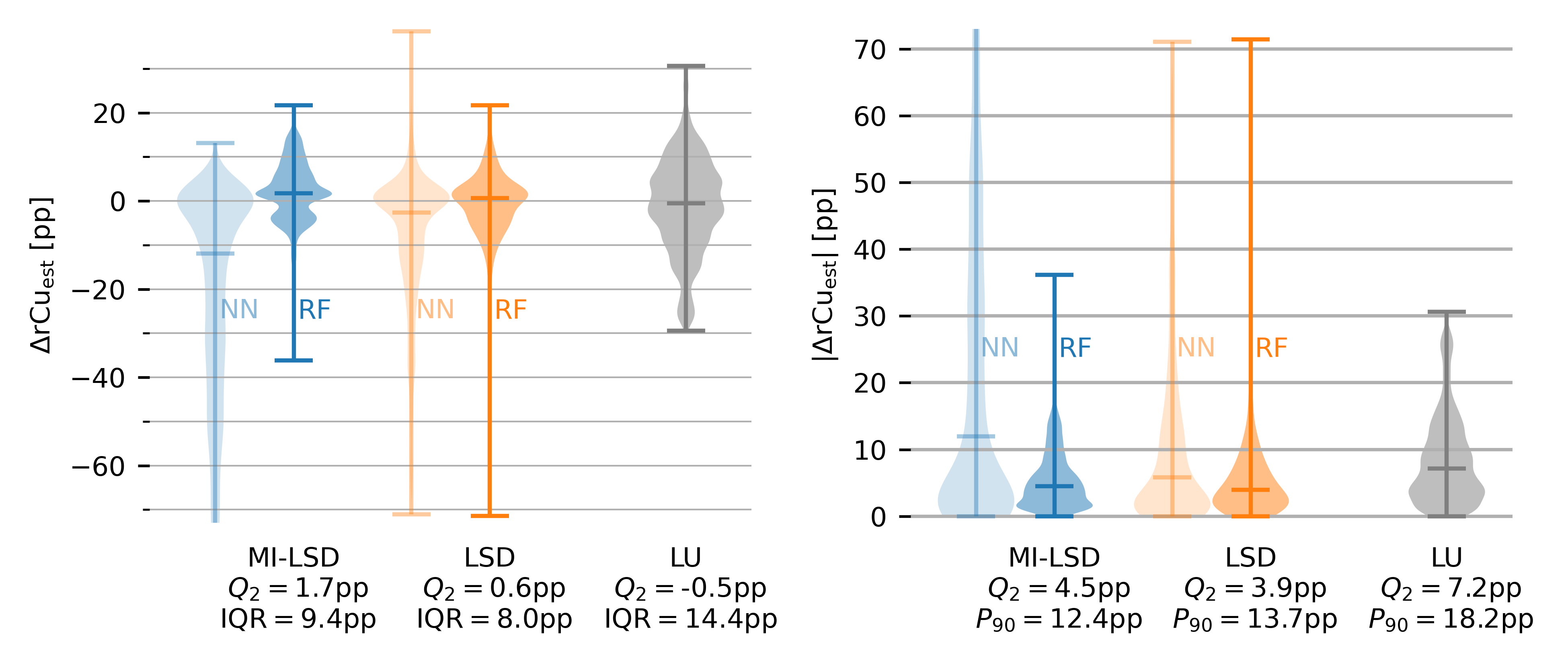}
\caption[Error distributions of the phantom test set C.]{\label{fig:errorsPhantomTestC}. Error distributions of the phantom test set C (cf.\ figure\,\ref{fig:phantoms}C). rCu estimation errors $\Delta \textrm{rCu}_\textrm{est}$ are shown left, their absolutes right. Blue shows the rCu estimators using multiple illumination learned spectral decoloring (MI-LSD), orange the estimators using learned spectral decoloring (LSD) and gray is the linear spectral unmixing (LU) reference. Medians $Q_2$ of the error distributions are shown, together with interquartile ranges (IQR) and 90 percentiles $P_{90}$. The feed forward neural network (NN) models performed similar to the random forest (RF) models for the LSD method but underperformed for MI-LSD.}
\end{figure}

The estimated rCu image examples from the test sets are shown for the RF models, because with the exception of the \emph{in silico} test set, NN models performed similarly or worse than RF models. For all estimated rCu images from all models, see the supplemental figures. The error distributions for phantom test set B are shown in figure\,\ref{fig:errorsPhantomTestB} and for phantom test set C in figure\,\ref{fig:errorsPhantomTestC}.
Descriptive statistics of the relative error distributions in the estimated rCu data are reported in Table\,1 for the two phantom test sets B and C.


\begin{table}[tbh]
\centering
\begin{tabular}{llllllllllll} 
\toprule
   &        & set & \multicolumn{4}{l}{$\Delta \textrm{rCu}_\textrm{est}$ [pp]} & \multicolumn{5}{l}{$|\Delta \textrm{rCu}_\textrm{est}|$ [pp]}  \\
   &        &     & mean  & $Q_1$ & $Q_2$ & $Q_3$                              & mean & $Q_1$ & $Q_2$ & $Q_3$ & $P_{90}$                       \\
RF & MI-LSD & B   & 0.6   & -2.7  & 1.1   & 3.4                                & 4.1  & 1.4   & 2.9   & 5.3   & 8.8                            \\
   &        & C   & 1.8   & -3.1  & 1.7   & 6.3                                & 5.6  & 2.1   & 4.5   & 7.9   & 12.4                           \\
   & LSD    & B   & -1.9  & -4.4  & 0.2   & 2.2                                & 5.2  & 1.5   & 3.3   & 6.2   & 10.7                           \\
   &        & C   & -2.8  & -5.3  & 0.6   & 2.6                                & 7.1  & 1.7   & 3.9   & 7.9   & 13.7                           \\
NN & MI-LSD & B   & -11.3 & -18.4 & -3.6  & 0.3                                & 12.8 & 1.3   & 5.3   & 18.4  & 36.2                           \\
   &        & C   & -21.0 & -38.2 & -12.0 & 0.1                                & 22.0 & 2.2   & 12.0  & 38.2  & 58.1                           \\
   & LSD    & B   & -3.1  & -5.9  & -0.3  & 1.8                                & 6.4  & 1.1   & 3.4   & 8.1   & 16.7                           \\
   &        & C   & -8.7  & -15.1 & -2.6  & 1.3                                & 11.4 & 1.7   & 5.8   & 15.7  & 32.6                           \\
LU &        & B   & -1.2  & -8.8  & 0.1   & 6.6                                & 8.2  & 4.0   & 7.4   & 11.1  & 15.2                           \\
   &        & C   & -1.0  & -8.0  & -0.5  & 6.3                                & 8.7  & 3.4   & 7.2   & 12.5  & 18.2                           \\
\bottomrule
\end{tabular}
\caption{Relative rCu estimation errors($\Delta \textrm{rCu}_\textrm{est}$) and absolute rCu estimation errors($|\Delta \textrm{rCu}_\textrm{est}|$) for the random forests (RF), neural networks (NN) and linear unmixing. Mean, median $Q_2$, 1st and 3rd quartiles $Q_1$ and $Q_3$, and the 90 percentile $P_{90}$ are listed for the phantom test sets B (transversal tubes) and C (longitudinal tubes).}
\end{table}

\section{Discussion}
The qualitative comparison of the absorbed energy spectra from the Monte Carlo simulations and the phantom PA signal spectra reveals a general agreement between the simulations and the phantom results. The existing variations between the normalized spectra of the two domains are likely due to discrepancies in the simulation e.g.\ the beam profiles and the the optical properties of the gel pad. The gel pad for example is currently simulated as water but also has some low-level scattering properties which was omitted in the simulation. Additionally, realistic laser noise was not simulated and the phantom positioning was only accurate to a millimeter. An acoustic forward simulation (e.g.\ using k-wave) was also not included in the simulation pipeline due to computational time constraints. While there are some acoustic artifacts (e.g.\ reflection artifacts) in the real PA image reconstructions, it is sensible to assume that they do not vary for different wavelength illumination, therefore their effect on spectral coloring should  be negligible. Variations in the Grüneisen parameter were also not part of the training set, even though it does vary significantly with rCu, because $\Gamma(c_\textrm{wb}(\textrm{NiSO}_4))\approx 0.21$ and $\Gamma(c_\textrm{wb}(\textrm{CuSO}_4))\approx 0.14$ at room temperature\cite{fonseca2017sulfates}. This results in a systematically higher SNR for low rCu -- an effect not present in sO$_2$\cite{savateeva2002optical}, which may explain why high rCu estimations are systematically worse in all of our phantom test sets. Laser noise levels are also wavelength dependent which is reinforced by the pulse energy correction -- e.g.\ resulting in a factor two SNR when measuring at 800\,nm compared to 680\,nm.

\begin{figure}[hbt]
\centering
\includegraphics{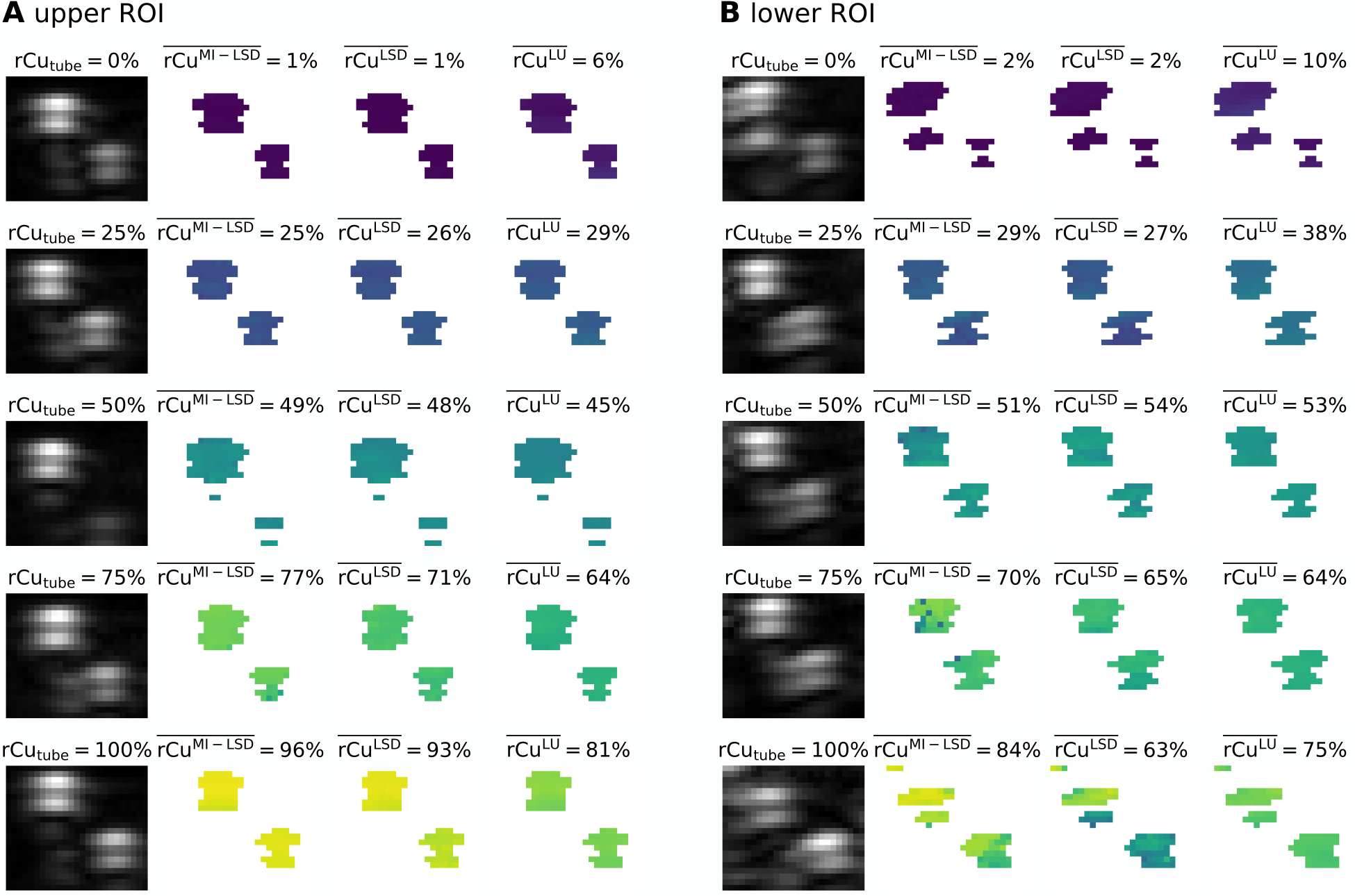}
\caption[Worst estimation example in Phantom test set B]{\label{fig:examplesPhantomTestB2} Worst estimation example in Phantom test set B: deep tubes (\textbf{B}) compared with more shallow tubes (\textbf{A}) in the same phantom with a svf = 0\,\%. Estimation results for MI-LSD, LSD and LU. Mean PA signal is shown left, with the ground truth rCu$_\textrm{tube}$ annotated. The mean rCu estimate $\overline{\textrm{rCu}}$ over the ROI is noted for the three estimators. Shallow tubes can be estimated very accurately while lower SNR in deep, high rCu tubes correlates to poor estimation accuracy.}
\end{figure}

Our MI-LSD method with RF estimators was highly accurate with median absolute estimation errors of only 2.9 and 4.5 percentage points in the two phantom test sets, respectively. Our NN models, however failed to give accurate estimates for MI-LSD. LSD estimates, using NN, were only improved over the LU reference and only in the phantom test set B. When testing on the out-of-distribution (OOD) test set C, our NN models showed no clear improvement over LU. This leads us to the initial conclusion that the overly complex NN models are prone to overfitting on the \emph{in silico} data, even when optimizing their hyperparameters with simple phantom data. The attempt to remedy this with dropout layers, lead to overall inaccurate estimations.

It is not surprising that the overall quantification performance was worse in deeper tubes. SNR in 8\,mm deep tubes was very low, i.e.\ the longer distance illuminations with 980\,nm light often yielded no detectable PA signal. This is due to background water absorption in combination with the high scattering, even when adding no sulfates to the background medium. We therefore also investigated omitting these higher wavelengths -- training and testing with fewer wavelengths from 680\,nm to 920\,nm, 20\,nm spaced. This yielded obviously worse model performance overall, which either suggests that (these) 13 wavelengths are insufficient for accurate estimation, or suggests that spectral coloring due to water absorption can be useful for a pixel wise correction of spectral coloring, as it can give implicit information on the optical path length. For further investigation it may be useful to add explicit information on the pixel position to the input features. It may also be interesting to perform similar experiments with a wider range of and/or more lower wavelength measurements, and then optimize the wavelength selection on these oversampled multispectral sequences. This was not done in this initial proof-of-concept work because it risks over-optimizing on unrealistic aspects of the rCu model (e.g.\ the difference in Grüneisen parameter of the two sulfate solutions) or setup specific aberrations (e.g.\ wavelength dependent SNR). Simulating more wavelengths also prolongs the already computationally expensive, 400 GPU days, simulation time for the necessary training data.

A final somewhat surprising observation was that estimation of both LSD and to a lesser extent MI-LSD is poorest in phantoms with no added sulfates in the background medium. Figure\,\ref{fig:examplesPhantomTestB2}B shows the worst estimation results in the lower tubes of phantom test set B -- combining three detrimental circumstances: (1) great depth (2) high rCu and (3) only spectral coloring of water. Though even in this worst case, MI-LSD is more accurate than the LSD or LU estimations.

One of the main shortcomings of the presented phantom validation is that it did not model melanin absorption in skin. Spectral coloring by melanin still causes large errors in standard of care pulse oximetry devices\cite{feiner_dark_2007, bickler2005effects} and needs to be addressed for qPAI. We were however not able to reproducibly include a skin mimicking layer with varying melanin absorption in our liquid phantoms -- future work will address this, by using solid, layered gel wax phantoms\cite{maneas2018gel}.

We showed a proof-of-concept setup with comparably poor image quality due to the US DAQ and transducer. An \emph{in vivo} applicable system should make use of state of the art US components and further engineering improvements to sensitivity and SNR as this currently limits the achievable estimation accuracy. Wavelength selection and illumination geometry are, so far, largely unoptimized.
Lastly, while the rCu model is a very useful tool for the investigation and thorough validation of a quantitative PA oximetry method, an explicit translation to actual sO$_2$ estimation \emph{in vivo} must be the next step. 

\section{Conclusion}
We presented MI-LSD, a quantitative photoacoustic oximetry method using multiple illuminations and machine learning; and presented a real-time MI PA imaging setup with a linear ultrasound transducer. We used this setup to image 115 phantom configurations by employing a highly reliable, reproducible and easily scalable phantom model.

MI-LSD with RFs was able to accurately and quickly estimate blood oxygen saturation modeled by copper and nickel sulfate. Compared to LU, MI-LSD approximately halved the absolute relative estimation error, achieving median absolute estimation errors of only 2.9 and 4.5 percentage points in our two phantom test sets, respectively.
To investigate such ML regression methods, thorough phantom validation is critical, as \emph{in silico} tests do not give sufficient data to validate a method, and \emph{in vivo} measurements lack a reliable ground truth. This is further illustrated by the fact that previously reported LSD NN models, which were only validated on \emph{in silico} data, slightly outperformed RF models on \emph{in silico} data (as was previously reported) but underperformed RF models in phantom tests while simply breaking on out-of-distribution (OOD) phantom data.

The results of this study give us a high degree of confidence that the domain gap from \emph{in silico} spectral decoloring to real data can be bridged using MI-LSD, paving the way to a clinical application of quantitative photoacoustic oximetry imaging.

\subsection*{Disclosures}
The authors have no relevant financial interests in this article and no potential conflicts of interest to disclose.

\subsection*{Acknowledgments}
This work has been funded by the Swiss National Science Foundation under project no.\,179038. Most calculations were performed on UBELIX, the HPC cluster at the University of Bern. We thank Michael Jaeger for his US data acquisition scripts, expertise and proof-reading; and Adrian Jenk at the Institute of Applied Physics mechanical workshop for his mechanical support. For their continuous support of the open source Medical imaging interaction toolkit (MITK) and the ippai python package, as well as for fruitful discussions we thank Janek Gröhl and the Photoacoustics team at the Computer Assisted Medical Interventions division, German Cancer Research Center, Heidelberg.

\subsection*{Author Contributions}
Conceptualization, T.K., M.F.; methodology, software, implementation, experiments, analysis, original draft preparation, T.K.; review and editing, T.K., M.F.; supervision, M.F.

\subsection*{Code, Data, and Materials Availability} 
The code for the methods as well as the experiments was implemented in python 3.7 and is fully open source, available at \href{https://github.com/thkirchner/PA-MI-LSD.git}{github:thkirchner/PA-MI-LSD}. All training, validation and test data sets generated in this work are openly available at \href{https://doi.org/10.5281/zenodo.4549631}{doi:10.5281/zenodo.4549631}. The raw Monte Carlo simulation results and raw PA scans are too large for upload -- 3\,TB -- but available from the authors upon reasonable request.

\bibliography{report}   
\bibliographystyle{spiejour}   

\listoffigures

\end{spacing}
\end{document}